\documentclass[aps,showpacs,preprint,groupedaddress,floatfix]{revtex4}
\usepackage{graphicx}
\usepackage{dcolumn}
\usepackage{bm}
\usepackage{amssymb}

\newcommand{\tfrac}[2]{{\textstyle {#1\over #2}}}

\newcommand{\bra}[1]{\left\langle #1 \right|}
\newcommand{\ket}[1]{\left| #1 \right\rangle}

\begin{document}

\title{Random-phase approximation based on relativistic 
     point-coupling models}
\author{T. Nik\v si\' c}
\author{D. Vretenar}

\affiliation{Physik-Department der Technischen Universit\"at M\"unchen, D-85748 Garching,
Germany, and \\
Physics Department, Faculty of Science, University of Zagreb, 
Croatia}
\author{P. Ring}
\affiliation{Physik-Department der Technischen Universit\"at M\"unchen, 
D-85748 Garching,
Germany}
\date{\today}

\begin{abstract}
The matrix equations of the random-phase approximation (RPA) 
are derived for the point-coupling Lagrangian of the relativistic 
mean-field (RMF) model. Fully consistent RMF plus (quasiparticle) RPA
illustrative calculations of the isoscalar monopole,
isovector dipole and isoscalar quadrupole response of spherical medium-heavy
and heavy nuclei, test the phenomenological effective interactions 
of the point-coupling RMF model. A comparison with experiment shows that 
the best point-coupling effective interactions accurately reproduce not 
only ground-state properties, but also data on excitation energies of 
giant resonances.
\end{abstract}
\pacs{21.60.-n, 21.30.Fe, 21.65.+f, 21.10.-k}
\maketitle

\section{\label{secI}Introduction}
The covariant self-consistent mean-field approach to the nuclear many-body
problem has reached a mature stage \cite{BHR.03,VALR.05}. Models based on 
the relativistic mean-field approximation are successfully employed 
in the description of structure phenomena not only in medium-heavy and
heavy stable nuclei, but also in regions of exotic nuclei far from the
line of $\beta$-stability and close to the nucleon drip-lines.

Most applications have been based on the finite-range meson-exchange 
representation of the relativistic mean-field (RMF) theory, in which 
the nucleus is described as a system of Dirac nucleons coupled to 
exchange mesons and the electromagnetic field through an effective 
Lagrangian. A medium dependence can be introduced either by
including non-linear meson self-interaction terms in the 
Lagrangian \cite{BB.77}, or by assuming an explicit density dependence 
for the meson-nucleon couplings. The former approach has been adopted
in the construction of several very successful phenomenological 
RMF interactions, for instance, the popular NL3~\cite{LKR.97} 
parameterization of the effective Lagrangian. In the latter case, 
the density dependence of the meson-nucleon vertex functions can 
be determined either from microscopic Dirac-Brueckner calculations 
in nuclear matter \cite{FLW.95,JL.98}, or it can be completely 
phenomenological \cite{TW.99,NVF.02}, with parameters adjusted to 
data on finite nuclei and empirical properties of symmetric and 
asymmetric nuclear matter. A series of recent studies has shown 
that, when compared with standard non-linear meson-exchange effective
Lagrangians, effective interactions with an explicit density 
dependence of the meson-nucleon couplings are more flexible and 
provide an improved description of asymmetric 
nuclear matter, neutron matter and finite nuclei far 
from stability. 

An alternative representation of the self-consistent relativistic 
mean-field approach to nuclear structure is formulated in terms 
of point-coupling (contact) nucleon-nucleon interactions. 
When applied in the description of finite nuclei,
the framework of relativistic mean-field point-coupling (RMF-PC) 
models \cite{MM.89,MNH.92,Hoch.94,FML.96,RF.97,BMM.02} produces results
that are comparable to those obtained in the meson-exchange 
picture. In principle, the point-coupling approach is more general 
and the interaction terms are not restricted by the constraints 
imposed by the finite range of meson exchange. Of course, also
in the case of contact interactions, medium effects can be taken 
into account by the inclusion of higher order interaction terms, 
for instance, six-nucleon vertices $(\bar\psi\psi)^3$, and
eight-nucleon vertices $(\bar\psi\psi)^4$ and
$[(\bar\psi\gamma_\mu\psi)(\bar\psi\gamma^\mu\psi)]^2$, or it 
can be encoded in the effective couplings, i.e. in the strength 
parameters of the interaction in the isoscalar and isovector channels. 
Several studies of the RMF-PC framework have been reported over 
the last ten years, but it is only recently that reliable and accurate 
phenomenological parameterizations have been adjusted and applied 
in the description of ground state properties of finite nuclei on
a quantitative level. In particular, based on an extensive  
multiparameter $\chi^2$ minimization procedure, in Ref.~\cite{BMM.02} 
B\"urvenich et al. have adjusted the PC-F1 set of coupling constants 
for an effective point-coupling Lagrangian with higher order interaction 
terms. The PC-F1 interaction has been tested in the calculation 
of ground state properties of a large number of spherical 
and deformed nuclei, and the results are on the level of the best 
meson-exchange effective interactions.  

Concepts of effective field theory and density functional theory
have been used to derive a microscopic relativistic point-coupling 
model of nuclear many-body dynamics constrained by
in-medium QCD sum rules and chiral symmetry \cite{FKV.03,FKV.04,VW.04}.
The effective Lagrangian is characterized by density-dependent 
coupling strengths, determined by
chiral one- and two-pion exchange and by QCD sum rule constraints
for the large isoscalar nucleon self-energies that arise through
changes of the quark condensate and the quark density
at finite baryon density. This approach has been tested in the analysis
of the equations of state for symmetric and asymmetric nuclear matter,
and of bulk and single-nucleon properties of finite nuclei.
In comparison with purely phenomenological mean-field approaches,
the built-in QCD constraints and the explicit treatment
of pion exchange restrict the freedom in adjusting
parameters and functional forms of density-dependent couplings.

A number of studies have shown that, both for finite-range meson-exchange
and for point-coupling mean-field models, the empirical
data set of ground-state properties of finite nuclei can
only determine six or seven parameters in the general expansion of
the effective Lagrangian in powers of the fields and their
derivatives \cite{FS.00b}. The influence of the adjustment procedure 
and of the choice of ground-state data on the properties and 
predictive power of the relativistic mean-field model with 
point-couplings has recently been investigated in Ref.~\cite{BMR.04}. 
While virtually all phenomenological relativistic effective interactions  
have been adjusted to empirical properties of symmetric and asymmetric 
nuclear matter, and to ground-state properties of a set of
spherical nuclei, in Ref. \cite{NVR.02} it has been shown that 
a comparison of relativistic RPA results on multipole
giant resonances with experimental excitation energies can provide 
additional constrains on the parameters that characterize the 
isoscalar and isovector channels of the effective interactions.
Data on giant resonances have been taken into account in the 
recent adjustment of a new improved relativistic 
mean-field effective interaction with explicit density 
dependence of the meson-nucleon couplings \cite{LNV.05}.

Relativistic RPA calculations have been performed since the 
early eighties, but it is only more recently that non-linear
meson self-interaction terms or density-dependent meson-nucleon 
couplings have been included in the RRPA 
framework \cite{Ma.97, VWR.00, NVR.02}. As in the case of 
ground-state properties, the inclusion of a medium dependence in 
the residual interaction is necessary for a quantitative 
description of collective excited states.   
Another essential feature of the RRPA is the fully consistent
treatment of the Dirac sea of negative energy states. In addition to
the usual particle-hole pairs, the RRPA configuration
space must also include pair-configurations built from 
positive-energy states occupied in the ground-state solution, 
and empty negative-energy states in the Dirac sea. These 
configurations ensure not only current conservation
and the decoupling of the spurious states~\cite{DF.90}, but also a
quantitative comparison with the experimental excitation energies of giant
resonances~\cite{Rin.01}. Collective excitations in open-shell nuclei 
can be analyzed with the relativistic quasiparticle 
random-phase approximation (RQRPA), which in Ref.~\cite{Paar.03} 
has been formulated in the canonical single-nucleon basis of the 
relativistic Hartree-Bogoliubov (RHB) model.

Some of the recent applications of the RRPA include studies of nuclear
compression modes~\cite{VWR.00, Ma.01, Pie.01}, of multipole giant resonances 
and low-lying collective states in spherical nuclei~\cite{Ma.02}, 
of the evolution of the low-lying isovector dipole response in nuclei 
with a large neutron excess~\cite{VPRLa.01, VPRLb.01}, and of the toroidal 
dipole response~\cite{VPNR.02}.
The RHB+RQRPA approach has been employed in the investigation of the multipole
response of weakly bound neutron-rich nuclei, and of spin-isospin
excitations in finite nuclei~\cite{Paar.03, Paar.04}. 

In this work we introduce the RPA based on the relativistic mean-field
framework with point-coupling interactions. Illustrative 
calculations of excitation energies of giant resonances in spherical 
nuclei will test the PC-F1 effective interaction \cite{BMM.02}.
The RRPA matrix equations are derived in Sec.~\ref{secII}. Isoscalar 
and isovector giant resonances in spherical nuclei are analyzed in 
Sec.~\ref{secIII}. The results are summarized in Sec.~\ref{secIV}.

\section{\label{secII}Random-phase 
approximation based on the point-coupling relativistic mean-field model}

The relativistic point-coupling Lagrangian is
built from basic densities and currents bilinear in the Dirac
spinor field $\psi$ of the nucleon:
\begin{equation}
  {\bar{\psi}} {\mathcal O}_\tau \Gamma  {\psi}
  \quad,\quad
  {\mathcal O}_\tau\in\{ {1},\tau_i\}
  \quad,\quad
  \Gamma\in\{1,\gamma_\mu,\gamma_5,\gamma_5\gamma_\mu,\sigma_{\mu\nu}\}\; .
\end{equation}
Here $\tau_i$ are the isospin Pauli matrices and
$\Gamma$ generically denotes the Dirac matrices.
The interaction terms of the Lagrangian are products of these
bilinears. Although a general effective Lagrangian
can be written as a power series in the currents 
${\bar{\psi}} {\mathcal O}_\tau \Gamma  {\psi}$ and their derivatives,
it is well known from numerous applications
of relativistic mean-field models that properties of symmetric and asymmetric
nuclear matter, as well as empirical ground state properties of finite
nuclei, constrain only the isoscalar-scalar (S), the isoscalar-vector (V),
the isovector-vector (TV), and to a certain extent the isovector-scalar (TS)
channels. Here we consider a model with four-, six-, and eight-fermion 
point couplings (contact interactions) \cite{BMM.02}, defined 
by the Lagrangian density:
\begin{equation}
\begin{array}{lcl}
  {\cal L} 
  & = & 
  {\cal L}^{\rm free} + {\cal L}^{\rm 4f} + {\cal L}^{\rm hot}
  + {\cal L}^{\rm der} + {\cal L}^{\rm em},
\\[12pt]
  {\cal L}^{\rm free} \hfill
  & = & 
  \bar\psi ({\rm i}\gamma_\mu\partial^\mu -m)\psi,
\\[6pt]
  {\cal L}^{\rm 4f} \hfill
  & = & 
  - \tfrac{1}{2}\, \alpha_{\rm S} (\bar\psi\psi)(\bar\psi\psi)
  - \tfrac{1}{2}\, 
    \alpha_{\rm V}(\bar\psi\gamma_\mu\psi)(\bar\psi\gamma^\mu\psi)
\\[6pt]
  & &
   - \tfrac{1}{2}\, \alpha_{\rm TS} (\bar\psi\vec\tau\psi) \cdot
   (\bar\psi\vec\tau\psi)
  - \tfrac{1}{2}\,  \alpha_{\rm TV} (\bar\psi\vec\tau\gamma_\mu\psi)
    \cdot (\bar\psi\vec\tau\gamma^\mu\psi),
\\[6pt]
  {\cal L}^{\rm hot} 
  & = &  
  - \tfrac{1}{3}\, \beta_{\rm S} (\bar\psi\psi)^3 - \tfrac{1}{4}\, 
    \gamma_{\rm S} (\bar\psi\psi)^4 - \tfrac{1}{4}\, \gamma_{\rm V} 
    [(\bar\psi\gamma_\mu\psi)(\bar\psi\gamma^\mu\psi)]^2,
\\[6pt]
  {\cal L}^{\rm der} 
  & = & 
  - \tfrac{1}{2}\,\delta_{\rm S}(\partial_\nu\bar\psi\psi)
    (\partial^\nu\bar\psi\psi)  
  - \tfrac{1}{2}\,  \delta_{\rm V} (\partial_\nu\bar\psi\gamma_\mu\psi)
    (\partial^\nu\bar\psi\gamma^\mu\psi)
\\[6pt]
  & &
   - \tfrac{1}{2}\, \delta_{\rm TS} (\partial_\nu\bar\psi\vec\tau\psi) \cdot
   (\partial^\nu\bar\psi\vec\tau\psi)
  - \tfrac{1}{2}\, \delta_{\rm TV} (\partial_\nu\bar\psi\vec\tau\gamma_\mu\psi)
    \cdot (\partial^\nu\bar\psi\vec\tau\gamma^\mu\psi),
\\[6pt]
  {\cal L}^{\rm em} 
  & = & 
  -  e A_\mu\bar\psi[(1-\tau_3)/2]\gamma^\mu\psi -  
    \tfrac{1}{4}\, F_{\mu\nu} F^{\mu\nu}.
\end{array}
\label{lagrangian}
\end{equation}
Vectors in isospin space are denoted by arrows, and bold-faced
symbols will indicate vectors in ordinary three-dimensional space.
In addition to the free nucleon Lagrangian $\mathcal{L}_{\rm free}$,
the four-fermion interaction terms contained in
$\mathcal{L}_{\rm 4f}$, and higher order terms in $\mathcal{L}_{\rm hot}$, 
when applied to finite nuclei the model
must include the coupling $\mathcal{L}_{\rm em}$
of the protons to the electromagnetic field $A^\mu$,
and derivative terms contained in $\mathcal{L}_{\rm der}$.
In the terms $\partial_\nu(\bar\psi \Gamma \psi)$ the derivative is
understood to act on both $\bar\psi$ and $\psi$. 
One could, of course, construct many more higher order interaction terms, 
or derivative terms of higher order, but in practice only a relatively 
small set of free parameters can be adjusted from the data set of 
ground state nuclear properties.

The Lagrangian is understood to be used in the mean-field approximation.
The single-nucleon Dirac equation is derived by the variation of the
Lagrangian (\ref{lagrangian}) with respect to $\bar{\psi}$
\begin{equation}
i\partial _{t}\psi _{i} =
\left\{\bm{\alpha}\lbrack-i\bm{\nabla}-
{\bm V(r},t{\bm )]+}V({\bm r},t)+
{\bm \beta }\big(m+S({\bm r},t)\big) \right\} \psi _{i} \;.
\label{dirac}
\end{equation}
The Dirac hamiltonian contains the scalar and vector potentials
\begin{equation}
S({\bm r},t) = \Sigma_S({\bm r},t) + \vec{\tau}\vec{\Sigma}_{TS}({\bm r},t)\;,
\end{equation}
\begin{equation}
V_{\mu}({\bm r},t)  = \Sigma^{\mu}({\bm r},t) 
+ \vec{\tau}\vec{\Sigma}^{\mu}_T({\bm r},t)\;,
\end{equation}
with the nucleon isoscalar-scalar, isovector-scalar, 
isoscalar-vector and isovector-vector self-energies 
defined by the following relations:
\begin{eqnarray}
\label{self1}
\label{self3}
   \Sigma_S & = & \alpha_S(\bar{\psi} \psi) + \beta_S (\bar{\psi} \psi)^2 + 
   \gamma_S(\bar{\psi} \psi)^3- \delta_S \Box (\bar{\psi} \psi) \; ,\\
\label{self4}
   \vec{\Sigma}_{TS} & = & \alpha_{TS} (\bar{\psi} \vec{\tau} \psi)
   -\delta_S \Box (\bar{\psi} \vec{\tau}\psi)\; , \\
\Sigma^{\mu} & = & \alpha_V (\bar{\psi} \gamma^\mu \psi) 
  +\gamma_V (\bar{\psi} \gamma^\alpha \psi)(\bar{\psi} \gamma_\alpha \psi)
   (\bar{\psi} \gamma^\mu \psi) - \delta_V \Box (\bar{\psi} \gamma^\mu \psi)     
   -eA^{\mu}\frac{1-\tau_3}{2} \; ,\\
\label{self2}
   \vec{\Sigma}^{\mu}_{T} & = & \alpha_{TV} (\bar{\psi} \vec{\tau} \gamma^\mu \psi) 
       - \delta_{TV} \Box (\bar{\psi} \vec{\tau}\gamma^\mu \psi)\; ,
\end{eqnarray}
respectively. The self-energies are determined by the corresponding 
local densities and currents calculated in the {\it no-sea}
approximation
\begin{eqnarray}
\rho _{S}({\bm r},t) &=&\sum\limits_{i=1}^{A}\bar{\psi}_{i}^{{}}({\bm r}%
,t)\psi _{i}^{{}}({\bm r},t)~,  \nonumber \\
\vec{\rho} _{TS}({\bm r},t) &=&\sum\limits_{i=1}^{A}
      \bar{\psi}_{i}^{{}}({\bm r},t)\vec{\tau}\psi _{i}^{{}}({\bm r},t)~,  \nonumber \\
j_{\mu }({\bm r},t) &=&\sum\limits_{i=1}^{A}\bar{\psi}_{i}^{{}}({\bm r}%
,t)\gamma _{\mu }\psi _{i}^{{}}({\bm r},t)~,  \nonumber \\
\vec{j}_{\mu }({\bm r},t) &=&\sum\limits_{i=1}^{A}\bar{\psi}_{i}^{{}}({\bm r}%
,t)\vec{\tau}\gamma _{\mu }\psi _{i}^{{}}({\bm r},t)~.  
\label{densities}
\end{eqnarray}
The summation runs over all A occupied states in the Fermi sea, i.e. only
occupied single-nucleon states with positive energy
explicitly contribute to the nucleon self-energies. Even though
the stationary solutions for the negative-energy states
do not contribute to the densities in the {\it no-sea} approximation,
their contribution is implicitly included in 
the time-evolution of the nuclear system~\cite{Rin.01,VBR.95}.

In an effective theory with the parameters of the Lagrangian
determined from a set of ground-state data, a large part
of vacuum polarization effects is already taken into
account in adjusting the parameters to experiment.
The stationary solutions of the relativistic mean-field equations
correspond to the ground-state of a nucleus. The Dirac spinors 
which determine the ground-state densities (i.e. positive-energy 
states) can be expanded, for instance, in terms of
vacuum solutions, which form a complete set of plane wave functions
in spinor space. This set is only complete, however, if in addition
to the positive-energy states, it also contains the states with
negative energy, in this case the Dirac sea of the vacuum.
Positive-energy solutions of the RMF equations in a finite nucleus
automatically contain vacuum components with negative energy. In the
same way, solutions that describe excited states, contain negative-energy 
components which correspond to the ground-state solution.

This is also true, in particular, for the solutions of the 
time-dependent problem. In the time-evolution of A 
nucleons in the effective mean-field potential, at each time $t$ the 
single-nucleon spinors $\psi _{i}(t)$ can be expanded in terms of 
the complete set of solutions of the stationary Dirac equation 
$\psi _{k}^{(0)}$. This means that at
each time $t$ one finds a {\it local} Fermi sea of $A$ time-dependent
spinors which, of course, contain components of negative-energy
solutions of the stationary Dirac equation. The states which form 
the {\it local} Dirac sea are orthogonal to the
{\it local} Fermi sea at each time. This is the meaning of the
{\it no-sea} approximation in the time-dependent problem.

The relativistic random-phase approximation (RRPA) equations can be derived from
the response of the density matrix to an external field that oscillates with a
small amplitude (for details see Refs.~\cite{Rin.01,NVR.02}).  
The matrix form of these equations reads
\begin{equation}
\left(
\begin{array}{cc}
A & B \\
B^{\ast } & A^{\ast }
\end{array}
\right) 
\left(
\begin{array}{c}
X_\nu \\
Y_\nu
\end{array} \right) = E_\nu \left(
\begin{array}{cc}
1 & 0 \\
0 & -1
\end{array} \right)
\left(
\begin{array}{c}
X_\nu \\
Y_\nu
\end{array} \right) \;,  
\label{RPAeq}
\end{equation}
where $E_\nu$ denotes the eigenfrequency, and $X_\nu$ and $Y_\nu$ are the
corresponding RPA amplitudes. The RRPA matrices A and B read
\begin{equation}
A =\left(
\begin{array}{cc}
(\epsilon _{p}-\epsilon _{h})\delta _{pp^{\prime }}\delta _{hh^{\prime }} &
\\
& (\epsilon _{\alpha }-\epsilon _{h})\delta _{\alpha \alpha ^{\prime
}}\delta _{hh^{\prime }}
\end{array}
\right) +\left(
\begin{array}{cc}
V_{ph^{\prime }hp^{\prime }} & V_{ph^{\prime }h\alpha ^{\prime }} \\
V_{\alpha h^{\prime }hp^{\prime }} & V_{\alpha h^{\prime }h\alpha ^{\prime }}
\end{array}
\right)   \label{RPAmatA} \end{equation}
and
\begin{equation}
B =\left(
\begin{array}{cc}
V_{pp^{\prime }hh^{\prime }} & V_{p\alpha ^{\prime }hh^{\prime }} \\
V_{\alpha p^{\prime }hh^{\prime }} & V_{\alpha \alpha ^{\prime }hh^{\prime }}
\end{array}
\right) \; .   \label{RPAmatB}
\end{equation}
The matrix elements of the residual interaction 
are derived from the Dirac hamiltonian of Eq. (\ref{dirac}):
\begin{equation}
V_{abcd} = \frac{\partial h_{ac}}{\partial \rho_{db}}  \;,
\end{equation}	
where the generic indices ($a,b,c,d,\ldots$) denote quantum numbers 
that specify the single-nucleon states $\{ \psi_a \}$. These belong to 
three distinct sets: the index $p$ (particle) denotes unoccupied states 
above the Fermi sea, the index $h$ (hole) is for occupied states in 
the Fermi sea, and with $\alpha$ we denote the unoccupied negative-energy 
states in the Dirac sea. 

Since the RRPA is derived in the small amplitude limit, 
the currents and densities can be expanded around their ground-state values
\begin{eqnarray}
\rho_{S}({\bm r},t) &=&\rho_{S}^{gs}({\bm r})+\delta \rho_{S}({\bm r},t)  \nonumber \\
\vec{\rho} _{TS}({\bm r},t) &=&\rho_{TS}^{gs}({\bm r}) 
    + \delta \vec{\rho}_{TS}({\bm r},t)\nonumber \\
j_{\mu }({\bm r},t) &=&\rho_{V}^{gs}({\bm r}) 
    + \delta j_{\mu }({\bm r},t)~,  \nonumber \\
\vec{j}_{\mu }({\bm r},t) &=& \rho_{TV}^{gs}({\bm r})
    + \delta \vec{j}_{\mu }({\bm r},t)\;.  
\label{expansion}
\end{eqnarray}
In this work we only consider spherical even-even nuclei. Because of 
time-reversal invariance, the spatial components of the currents vanish in
the nuclear ground state. Furthermore, charge conservation 
implies that only the 3-component of the isovector scalar and vector 
densities contributes in the ground state.
The individual contribution of each field to $V_{abcd}$ can now be obtained
by inserting the expansions Eq. (\ref{expansion}) in the matrix element 
of the Dirac Hamiltonian Eq. (\ref{dirac}):
\begin{eqnarray}
V_{abcd}^S &=& \int{\psi_a^\dagger\beta\psi_c \left( \alpha_S 
  + 2\beta_S\rho_S^{gs} + 3\gamma_S{\rho_S^{gs}}^2 + \delta_S\Delta \right)
    \psi_b^\dagger\beta\psi_d\; d^3r}\;, \nonumber \\
V_{abcd}^{TS}&=& \int{\psi_a^\dagger\beta\vec{\tau}\psi_c \left( \alpha_{TS} 
  + \delta_{TS}\Delta \right) \psi_b^\dagger\beta\vec{\tau}\psi_d\; d^3r}\;,
    \nonumber \\
V_{abcd}^{V}&=& \int{\psi_a^\dagger\psi_c \left( \alpha_{V} 
  + 3\gamma_V{\rho_V^{gs}}^2 + \delta_V\Delta \right) 
  \psi_b^\dagger\psi_d\; d^3r} \nonumber \\ &-&
  \int{\psi_a^\dagger\bm{\alpha}\psi_c \left( \alpha_{V} 
  + \gamma_V{\rho_V^{gs}}^2 + \delta_V\Delta \right) 
  \psi_b^\dagger\bm{\alpha}\psi_d\; d^3r}\;, \nonumber \\
V_{abcd}^{TV}&=& \int{\psi_a^\dagger\beta\vec{\tau}\gamma_{\mu}\psi_c 
 \left( \alpha_{TV} + \delta_{TV}\Delta \right) 
 \psi_b^\dagger\beta\vec{\tau}\gamma^{\mu}\psi_d\; d^3r}\;.
\label{Vvecvec}    
\end{eqnarray}
  
For open-shell nuclei calculations are performed in the framework 
of the fully self-consistent RHB plus relativistic QRPA model. The
RHB represents a relativistic extension of the Hartree-Fock-Bogoliubov
model, and it provides a unified description of particle-hole 
($ph$) and particle-particle ($pp$) correlations.
In most applications of the RHB model \cite{VALR.05}
the pairing part of the well known and very successful 
Gogny force~\cite{BGG.84} has be employed in the
$pp$ channel.
\begin{equation}
V^{pp}(1,2)~=~\sum_{i=1,2}e^{-((\mathbf{r}_{1}-\mathbf{r}_{2})/{\mu_{i}})^{2}%
}\,(W _{i}~+~B_{i}P^{\sigma}-H_{i}P^{\tau}-M_{i}P^{\sigma}P^{\tau})\;,
\end{equation}
with the set D1S \cite{BGG.91} for the parameters $\mu_{i}$,
$W_{i}$, $B_{i}$, $H_{i}$, and $M_{i}$ $(i=1,2)$. This force has
been very carefully adjusted to the pairing properties of finite
nuclei all over the periodic table. In particular, the basic
advantage of the Gogny force is the finite range, which
automatically guarantees a proper cut-off in momentum space.

In Ref.~\cite{Paar.03} the RQRPA has been formulated 
in the canonical single-nucleon basis of the RHB model. 
By definition, the canonical basis diagonalizes the density 
matrix and it is always localized. It describes both the bound 
states and the positive-energy single-particle
continuum. This particular representation of the RQRPA is
very convenient because, in order to describe transitions to
low-lying excited states in weakly-bound nuclei, the
two-quasiparticle configuration space must include states with both nucleons
in the discrete bound levels, states with one nucleon in a bound level and
one nucleon in the continuum, and also states with both nucleons in the
continuum. In addition, the full RQRPA equations are rather complicated 
and it is considerably simpler to solve these equations in the canonical 
basis where the RHB wave functions take a simple BCS-form. In this case 
one needs only the matrix elements of the interactions in the 
$ph$ and $pp$ channels, and certain combinations of the occupation 
factors of canonical states.
The relativistic QRPA of Ref.~\cite{Paar.03} is fully self-consistent.
For the interaction in the particle-hole channel effective Lagrangians with
nonlinear meson self-interactions or nucleon point couplings are used, 
and pairing correlations are described by the pairing part of the finite 
range Gogny interaction. Both in the $ph$ and $pp$ channels, 
the same interactions are used in the RHB
equations that determine the canonical quasiparticle basis, and in the
matrix equations of the RQRPA. The RQRPA configuration space 
includes also the Dirac sea of negative energy states.

In the next section we will present illustrative relativistic
RPA/QRPA calculations of the multipole response of spherical nuclei. 
For the multipole operator $\hat{Q}_{\lambda \mu}$,
the response function $R(E)$ is defined
\begin{equation}
R(E) = \sum_f B(0_i\to \lambda_f )\frac{\Gamma/ 2\pi}{(E-E_f)^2+(\Gamma/2)^2}\;,
\label{response}
\end{equation}
where $\Gamma$ is the width of the Lorentzian distribution, and
\begin{equation}
B(0_i\to \lambda_f )=\left|\bra{\lambda_f}| \hat{Q}_{\lambda}| \ket{0_i} \right|^2\;.
\label{strength}
\end{equation}
For all calculations in this work the discrete spectrum of the RRPA states 
is folded with the Lorentzian of Eq. (\ref{response}), with the 
width $\Gamma = 1$ MeV.	
\section{\label{secIII}Multipole giant resonances}	     

We have performed fully consistent relativistic RPA/QRPA calculations 
of isoscalar monopole, isovector dipole, and isoscalar quadrupole 
giant resonances in spherical nuclei. The interaction in the particle-hole
channel is determined by the effective point-coupling Lagrangian 
Eq.~(\ref{lagrangian}), and pairing correlations are described by 
the pairing part of the finite range Gogny interaction D1S \cite{BGG.91}. 
The R(Q)RPA configuration space includes the
Dirac sea of negative energy states. Both in the particle-hole and
particle-particle channels, the same interactions are used in the 
calculation of the ground state and in the matrix equations of the R(Q)RPA.

The point-coupling Lagrangian Eq.~(\ref{lagrangian}) contains 11 adjustable 
coupling constants. The PC-F1 effective interaction, adjusted in 
Ref.~\cite{BMM.02}, corresponds to a restricted set of 9 coupling parameters
and does not include the isovector-scalar channel. The parameters have 
been determined in a $\chi^2$-minimization procedure, adjusted to 
ground-state observables (binding energy, charge radius, diffraction radius and
surface thickness) of 17 spherical nuclei. The resulting parameters of the
PC-F1 effective interaction are displayed in Tab.~\ref{tab0}. This interaction 
has been tested in the analysis of the equation of state of symmetric nuclear
matter and neutron matter, binding energies and form-factor- and 
shell-structure-related ground-state properties of several isotopic and 
isotonic chains, including superheavy nuclei with known experimental masses, 
and of the fission barrier in $^{240}$Pu. The comparison with data has shown
that the RMF-PC model with the PC-F1 interaction can compete with the 
best phenomenological finite-range meson-exchange interactions. It should
be noted, however, that PC-F1 exhibits a relatively large volume asymmetry 
at saturation $a_4 \approx 38$ MeV, resulting in a very stiff equation of
state for neutron matter, and too large values for the neutron skin in 
finite nuclei. The most recent meson-exchange RMF effective forces, on the 
other hand, include an explicit medium dependence 
both in the isoscalar and isovector channels~\cite{TW.99,NVF.02}, and thus
provide an improved description of asymmetric nuclear matter and
neutron matter, and realistic values of the neutron skin.

In the following examples we will test the PC-F1 interaction in the calculation
of excitation energies of giant resonances in spherical nuclei. We will 
also try to determine whether R(Q)RPA calculations of excited states 
can be used to discriminate between different point-coupling interactions,    
or even place additional constraints on the parameters of the 
effective interactions~\cite{NVR.02}.  
\subsection{\label{subIIIa}The isoscalar giant monopole resonance}   

The isoscalar giant monopole resonance (ISGMR) represents the simplest mode of
collective oscillations in finite nuclei (the breathing mode), 
and provides valuable information on the nuclear matter incompressibility.
The range of values of the nuclear matter compression modulus $K_\infty$ 
can be best determined by comparing results of
fully consistent microscopic calculation of both ground state
properties and the ISGMR excitation energies in spherical nuclei with data.
Moreover, since $K_{\infty}$ determines
bulk properties of nuclei and, on the other hand,
the GMR excitation energies depend also on the surface compressibility,
measurements and microscopic calculations of GMR in
heavy spherical nuclei should, in principle, provide a more reliable
estimate of the nuclear matter incompressibility \cite{Bla.80, BBDG.95}. 
A recent relativistic mean-field plus R(Q)RPA analysis of the ISGMR,
based on effective Lagrangians with density-dependent meson-nucleon
vertex functions, has shown that the nuclear matter compression
modulus of effective interactions based on the relativistic
mean-field approximation should be restricted to a rather narrow interval
$K_{\infty}\approx 250 - 270$ MeV \cite{VNR.03}.

Although the point-coupling PC-F1 interaction has been 
adjusted to ground-state data only, its compression modulus 
$K_\infty = 270$ MeV is within the range predicted by 
the meson-exchange models. The latter, however, 
tend to underestimate the surface thickness of finite nuclei \cite{BMM.02}.
Since the excitation energies of the ISGMR generally depend 
also on the surface compressibility, point-coupling 
and meson-exchange effective interactions with nearly identical values of 
the nuclear matter compression modulus, do not necessarily predict 
identical GMR excitation energies, especially in lighter nuclei 
in which surface effects are more pronounced. 

\begin{figure}
\includegraphics[scale=0.6,clip]{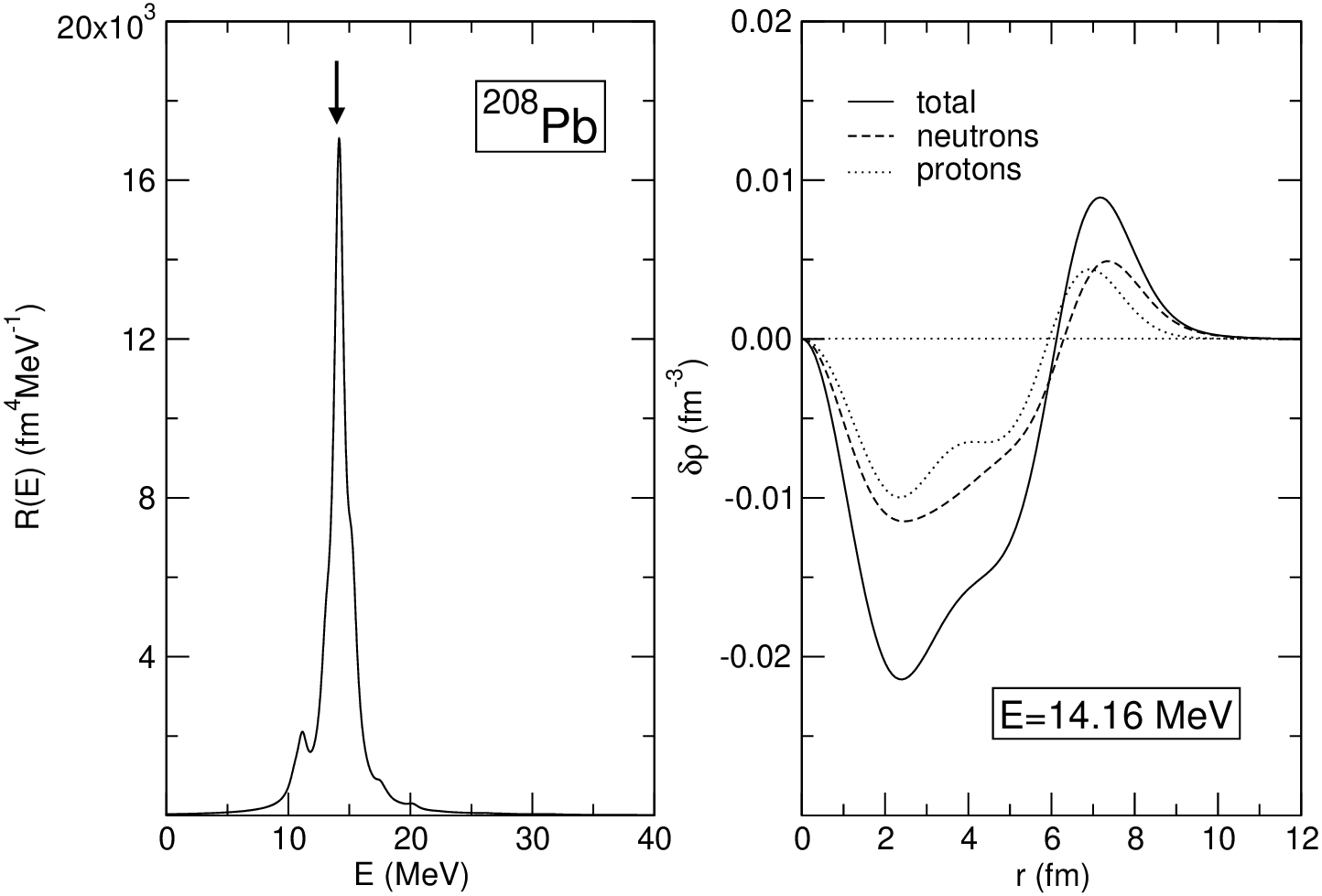}
\caption{The isoscalar monopole strength distribution (left panel) and the
transition densities (right panel) in $^{208}$Pb, calculated with the PC-F1
interaction. The experimental excitation energy of the ISGMR is denoted by the
arrow. The neutron, proton and total isoscalar transition densities correspond
to the peak at $E=14.16$ MeV.}
\label{figA}
\end{figure}

In Fig.~\ref{figA} we display the isoscalar monopole strength distribution 
in $^{208}$Pb, calculated in the relativistic RPA with the PC-F1 effective
interaction. For the ISGMR peak at $E=14.16$ MeV excitation energy, in the 
right panel we plot the corresponding proton, neutron and total isoscalar
transition densities. The node at the surface is characteristic for the 
the breathing mode of oscillations. The calculated peak energy is in 
excellent agreement with the newest data on the ISGMR centroid energy 
$m_1 /m_0 = 13.96 \pm 0.20$ MeV from Ref.~\cite{Young2} (denoted by the arrow in
Fig.~\ref{figA}). In Table \ref{tab1} we have 
also compared the R(Q)RPA $m_1 /m_0$ centroids with very recent data 
on medium-heavy nuclei \cite{Young1,Young2,Young3}. It appears that 
the excitation energies predicted by the PC-F1 effective interaction are
systematically somewhat higher than the experimental ISGMR's, indicating 
that the nuclear matter compression modulus of a relativistic point-coupling 
model should probably be closer to $K_{\infty}\approx 250$ MeV. This would
be in agreement with the modern density-dependent meson-exchange effective
interactions: DD-ME1 with $K_\infty = 245$ MeV \cite{NVF.02}, and 
DD-ME2 with $K_\infty = 251$ MeV \cite{LNV.05}.

\subsection{\label{subIIIb}The isovector giant dipole resonance}   

The isovector giant dipole resonance (IVGDR), calculated in the R(Q)RPA, is
predominantly determined by the isovector channel of the effective interaction.
In particular, the position of the IVGDR is directly related to the nuclear
matter asymmetry energy. This quantity can be expanded in a Taylor series in
$\rho$~\cite{Lee.98}
\begin{equation}
S_2(\rho) = a_4 + \frac{p_0}{\rho_{sat}^2}(\rho - \rho_{sat})+
     \frac{\Delta K}{18\rho_{sat}^2}(\rho - \rho_{sat})^2+ \cdots \;.
\end{equation}
The value of the asymmetry energy at the saturation density (volume asymmetry)
is denoted by $a_4$, the parameter $p_0$ defines the linear density dependence
of the asymmetry energy, and $\Delta K$ is the correction to the
incompressibility. The asymmetry energy directly determines 
the difference $r_n -r_p$ between the radii of the neutron and proton 
ground-state density distributions. In a recent study that has analyzed 
available data on neutron radii and the excitation energies of 
the IVGDR in the framework of
the density-dependent meson-exchange RMF models, 
the volume asymmetry of RMF effective interactions has been constrained to
the interval: $32$ MeV $\le a_4 \le$ $36$ MeV~\cite{VNR.03}. For the
PC-F1 effective interaction the volume asymmetry is somewhat larger:
$a_4 = 37.8$ MeV. This is also the case for older 
meson-exchange RMF forces with non-linear meson self-interaction terms, 
and is due to the fact that the isovector channel of these interactions 
is basically parameterized by a single constant: $\alpha_{\rm TV}$ in the 
point-coupling version, or the $\rho$-meson coupling $g_\rho$ in the 
meson-exchange models. With a single parameter in the isovector channel, 
it is simply not possible to simultaneously lower $a_4$ 
to its empirical value and reproduce the masses of $N \neq Z$ nuclei.
This only becomes possible if a density dependence is included in the 
isovector channel, as it is done in modern density-dependent meson-exchange 
RMF forces~\cite{TW.99, NVF.02,LNV.05}. 

\begin{figure}
\includegraphics[scale=0.6,clip]{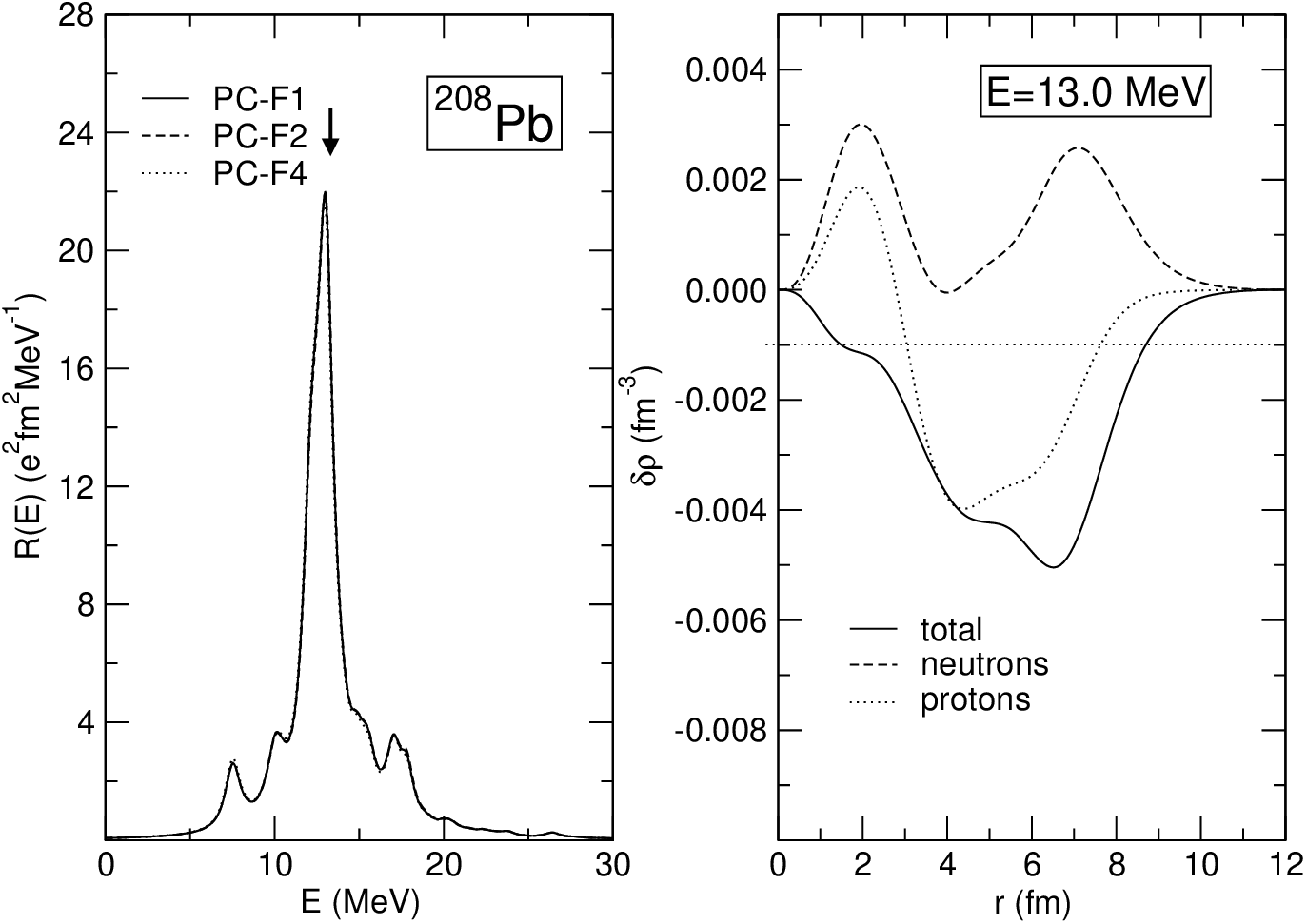}
\caption{The isovector dipole strength distribution (left panel) in $^{208}$Pb, 
calculated with the PC-F1, PC-F2 and PC-F4 effective interactions. 
The experimental
excitation energy of the IVGDR is denoted by the arrow. 
In the right panel we plot
the neutron, proton and total isovector transition densities for
the peak at $E=13.0$ MeV, and calculated with PC-F1.}
\label{figB}
\end{figure}

\begin{figure}
\includegraphics[scale=0.6,clip]{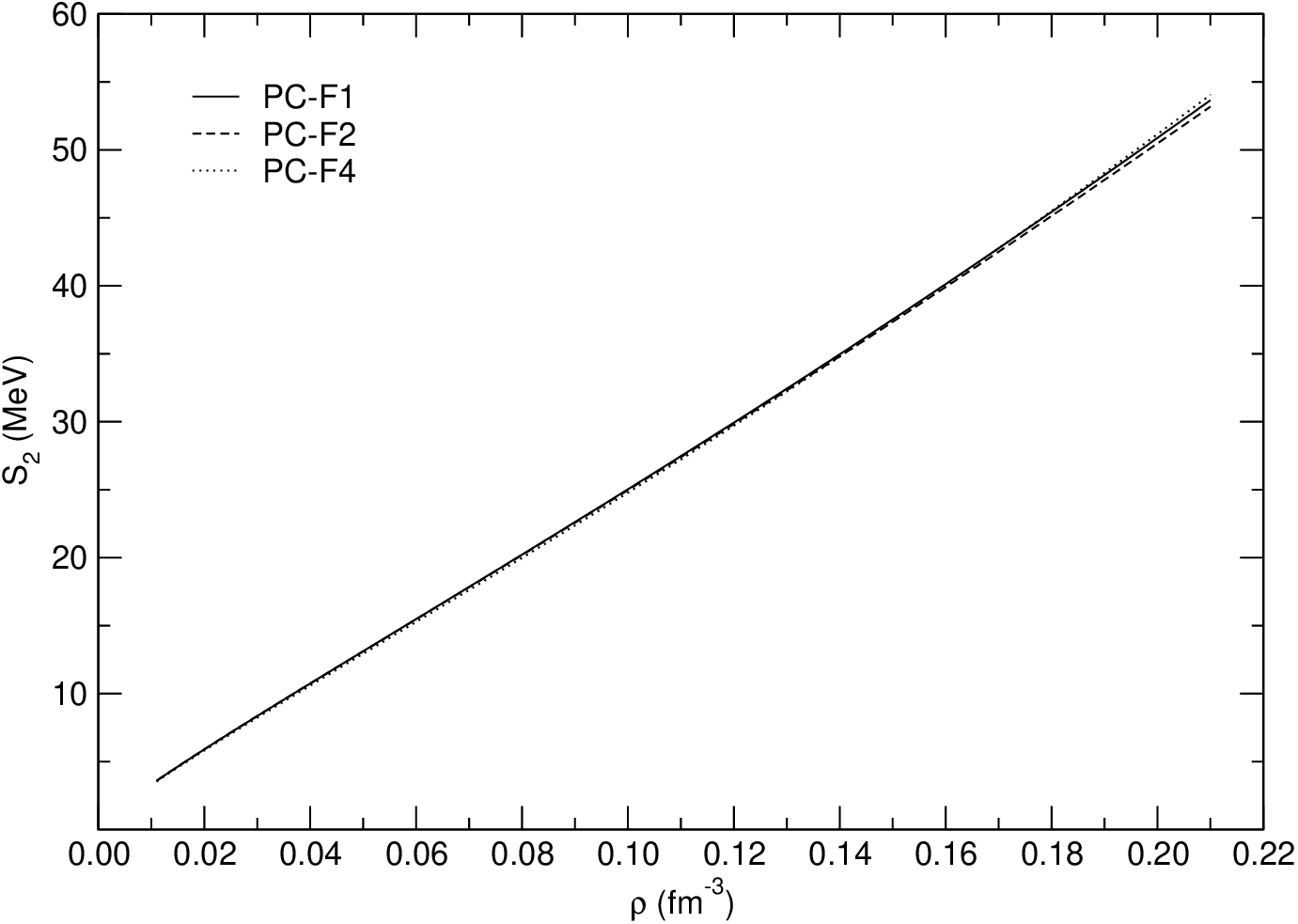}
\caption{The nuclear matter asymmetry energy as function of the nucleon density,
calculated with the PC-F1, PC-F2 and PC-F4 effective interactions.}
\label{figC}
\end{figure}

For the point-coupling effective Lagrangian Eq.~(\ref{lagrangian}), in 
Ref.~\cite{BMM.02} it has been shown that the strength parameter 
$\delta_{\rm TV}$ of the derivative term in the isovector channel cannot be 
determined from ground-state properties of finite nuclei. The isovector 
channel of the point-coupling Lagrangian was also investigated by the 
inclusion of the isovector-scalar terms. Two additional
interactions were generated: PC-F2 which includes only the linear
isovector-scalar term with the coupling constant $\alpha_{\rm TS}$, 
and PC-F4 which contains both the linear and derivative isovector-scalar 
terms, with the corresponding parameters $\alpha_{\rm TS}$ and 
$\delta_{\rm TS}$. In comparison to the PC-F1 interaction,
however, the $\chi^2$ for the extended sets PC-F2 and PC-F4 was 
reduced by less then $1\%  
$, and it was concluded that these 
extensions are not well determined by the ground-state data included 
in the fit. Since properties of isovector collective 
modes could, in principle, provide additional information on the isovector
channel of the effective interaction, it is interesting to compare the
isovector dipole strength distributions calculated with PC-F1, PC-F2, and PC-F4.
For $^{208}$Pb the resulting curves, shown
in the left panel of Fig.~\ref{figB}, are practically indistinguishable. 
This is simply because the corresponding asymmetry energies begin 
to differ only at densities high above the saturation density 
(see Fig.~\ref{figC}). On the other hand, since the IVGDR corresponds to 
a predominantly surface mode of oscillations, the density region which 
determines this resonance is located below saturation density. This is 
illustrated in the right panel of Fig.~\ref{figB}, where we plot
the neutron, proton, and total isovector transition densities for
the peak at $E= 13.0$ MeV, calculated with the PC-F1 interaction. 
For the point-coupling Lagrangian, this means that properties of the 
IVGDR do not determine more precisely the couplings 
$\alpha_{\rm TS}$ and $\delta_{\rm TS}$. 

We note that the PC-F1 interaction predicts the excitation energy of the IVGDR
in $^{208}$Pb at $E=13.0$ MeV, which is below the experimental value 
$E=13.3\pm 0.1$ MeV~\cite{Rit.93}. This is also the case for the lighter 
nuclei: $^{90}$Zr, $^{116}$Sn, $^{118}$Sn, $^{120}$Sn, and $^{124}$Sn, 
for which in Table \ref{tab2} we compare the R(Q)RPA IVGDR excitation 
energies with data \cite{BF.75}. The fact that effective interactions 
with large volume asymmetry underestimate the energy of the IVGDR has 
already been demonstrated in two recent studies of the isovector dipole 
response performed with non-relativistic 
and relativistic RPA \cite{Rei.99, NVR.02}.

\subsection{\label{subIIIc}The isoscalar giant quadrupole resonance}   

In nonrelativistic RPA calculations, the excitation energy of the isoscalar
giant quadrupole resonance (ISGQR) can be directly related to the nucleon
effective mass $m^*$ associated with a given interaction. 
For Skyrme-type effective interactions, in particular, the excitation energy of
the ISGQR exhibits a linear dependence on $m^*$. The larger the effective mass, i.e.,
the higher the density of states around the Fermi surface, the lower is the
calculated ISGQR excitation energy. Calculations of both ground-state
properties and ISGQR excitation energies in spherical nuclei, 
constrain the effective mass for Skyrme-type 
interactions to the interval: $m^* /m = 0.8 \pm 0.1$~\cite{Rei.99}.

The situation is slightly more complicated in the relativistic framework, 
because one finds several different quantities denoted as the "effective mass".
The quantity which is most often used to characterize an effective 
interaction is the Dirac mass
\begin{equation} 
m_D = m+ S(\bf{r})\;,
\label{dirac_mass}
\end{equation}
where m is the nucleon mass and $S(\bf{r})$ is the scalar nucleon self-energy.
The concept of the effective mass in the relativistic framework has been
extensively analyzed in Refs.~\cite{JM.89, JM.90}. In particular, it has been
shown that one must not identify the Dirac mass with the effective mass of
the nonrelativistic mean-field models. Instead, the quantity which should be
compared with the empirical effective mass derived from nonrelativistic
analyses of scattering and bound state data is given by
\begin{equation}
m^* = m - V(\bf{r})\;,
\label{eff_mass}
\end{equation}
where $V(\bf{r})$ denotes the time-like component of the vector self-energy.
However, both $m_D$ and $m^*$ are essentially determined by: 
(i) the empirical spin-orbit splittings in finite nuclei, and (ii)
the binding energy at saturation density in nuclear matter. They place the
following constraints on the effective masses: $0.55 m \le m_D \le 0.6 m$ and 
$0.64 m \le m^* \le 0.67 m$. In comparison to the nonrelativistic
self-consistent mean-field models, the allowed values for the 
relativistic $m^*$ are rather low, resulting in a smaller density of
states around the Fermi surface. Moreover, the allowed interval of $m^*$ values
is so narrow, that there is no room for any significant enhancement of the
single-nucleon level densities in the framework of the standard 
phenomenological RMF models \cite{VNR.02}.

\begin{figure}
\includegraphics[scale=0.6,clip]{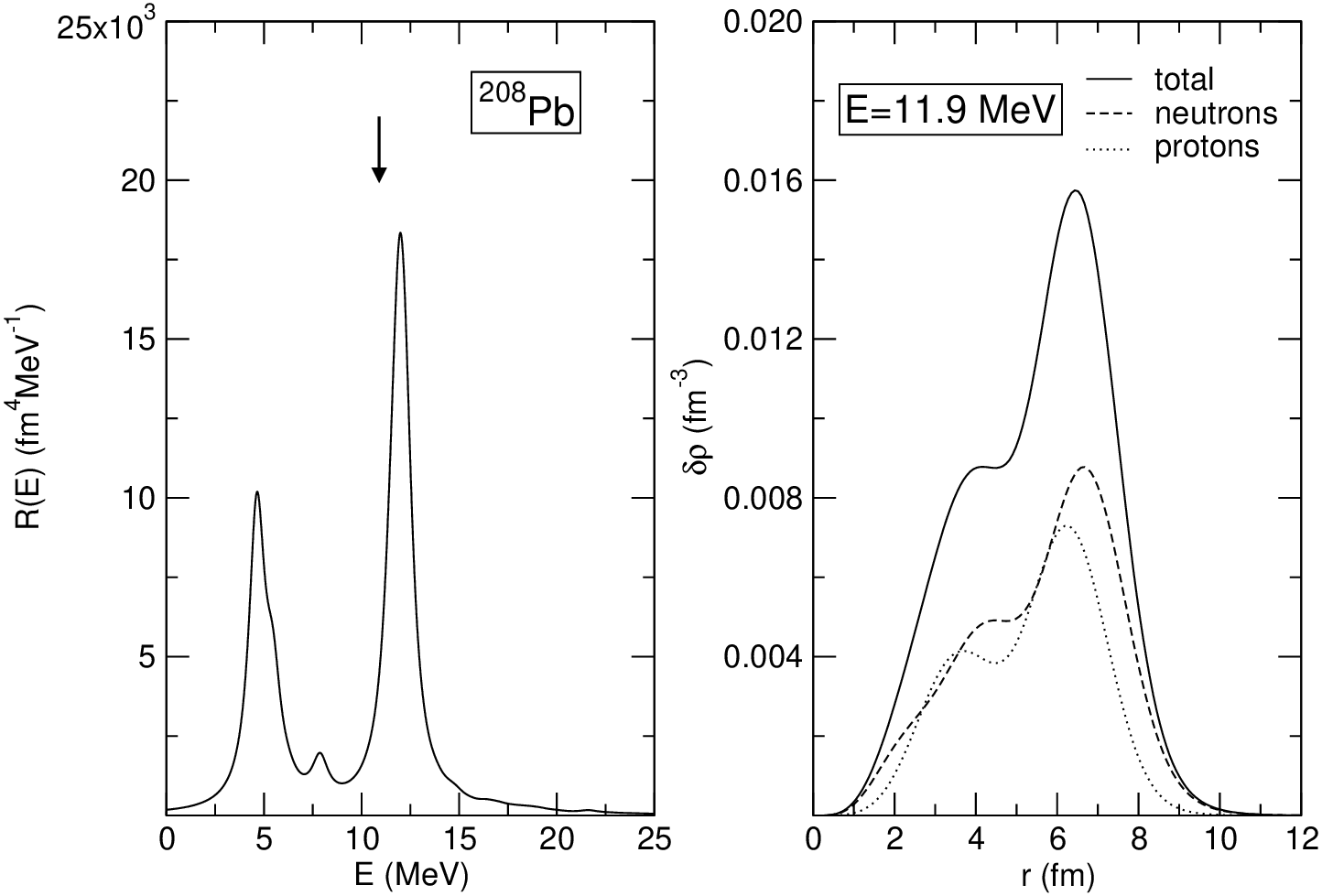}
\caption{The isoscalar quadrupole strength distribution (left panel) and
the transition densities (right panel) for $^{208}$Pb,
calculated with the PC-F1 effective interaction. 
The experimental excitation energy of the ISGQR is denoted by the arrow. 
The neutron, proton and total
isoscalar transition densities correspond to the ISGQR peak at $E=11.9$ MeV.}
\label{figD}
\end{figure}

These arguments are also valid for point-coupling RMF models.
Specifically, for the PC-F1 effective interaction the Dirac mass 
$m_D = 0.61 m$, the effective mass $m^* = 0.69 m$, and therefore one should 
not expect PC-F1 to accurately reproduce data on the ISGQR. 
In the left panel of Fig.~\ref{figD} we plot the RRPA
isoscalar quadrupole strength distribution in $^{208}$Pb, calculated with the
PC-F1 interaction. The experimental excitation energy of the ISGQR 
($10.89 \pm 0.3$ MeV \cite{Young2}) is denoted by the arrow. 
Because of the low nucleon effective mass, the calculated excitation energy
of the ISGQR is above the corresponding 
experimental value. Similar results are also obtained for lighter 
spherical nuclei. In Table \ref{tab3} we display a comparison between 
the experimental excitation energies 
and the PC-F1 predictions for the 
location of the ISGQR in $^{90}$Zr, $^{116}$Sn, $^{112}$Sn, $^{124}$Sn, 
$^{144}$Sm, and $^{208}$Pb. For all these nuclei the calculated ISGQR 
excitation energy is more than 1 MeV above the experimental centroids 
$m_1 /m_0$.
\section{\label{secIV}Summary and outlook}		     

During the last decade standard meson-exchange RMF models, with either
non-linear meson self-interaction terms, or with
density-dependent meson-nucleon vertex functions, have been very 
successfully applied in the description of a variety of
nuclear structure phenomena. 
However, the explicit inclusion of the meson degrees of freedom, 
in particular of the fictitious $\sigma$-meson, places physical 
constraints on the model parameters,
thereby reducing the predictive power of the model. The limitations of the
meson-exchange representation of the RMF theory are especially pronounced 
in the description of surface properties of finite nuclei. 
Virtually all meson-exchange RMF effective interactions, which 
otherwise accurately reproduce data on bulk nuclear
properties and giant resonances, underestimate the empirical  
surface thickness. This does not seem to be the case for the self-consistent
point-coupling RMF models, which therefore represent an interesting 
alternative to the standard meson-exchange 
picture of the effective nuclear interaction. 

Self-consistent point-coupling RMF models have recently attracted 
considerable interest. For the phenomenological models,
in particular, it has been shown that the new PC-F1 effective interaction 
reproduces data with a quality comparable to that of standard meson-exchange
forces. However, all calculations performed so far have only considered
ground-state nuclear properties. It is, therefore, important to develop 
a consistent microscopic framework, based on the point-coupling RMF 
effective Lagrangian, in which dynamical properties and excited states 
can be investigated. 

In this work the matrix equations of the random-phase approximation (RPA) 
have been derived for the point-coupling Lagrangian of the (RMF) model. 
Fully consistent RMF plus RPA, and RHB plus QRPA 
illustrative calculations of the isoscalar monopole,
isovector dipole and isoscalar quadrupole response of spherical medium-heavy
and heavy nuclei have been performed. A comparison with experiment has shown 
that the best point-coupling effective interactions, and in particular 
PC-F1, accurately reproduce not only ground-state properties, but also data 
on excitation energies of giant resonances.
We have also investigated the possibility to determine the parameters of the 
isovector-scalar channel from R(Q)RPA calculations of the isovector dipole
response. This is really not feasible because
the isovector-scalar terms influence the symmetry energy only for nucleon
densities well above the saturation density, whereas the density region
characteristic for the IVGDR is located below the saturation density.

The R(Q)RPA based on the point-coupling RMF models presents an 
important addition to the theoretical tools that are employed in 
description of the nuclear many-body problem. Future applications will
include studies of the multipole response of exotic nuclei far from
the valley of $\beta$-stability. On a more microscopic level, the 
R(Q)RPA will be used to investigate dynamical properties predicted by
the recently introduced relativistic point-coupling model constrained by
in-medium QCD sum rules and chiral symmetry \cite{FKV.03,FKV.04,VW.04}.
\bigskip 
\noindent
\bigskip \bigskip

\bigskip \bigskip
\leftline{\bf ACKNOWLEDGMENTS}
This work has been supported in part by the Bundesministerium
f\"ur Bildung und Forschung - project 06 MT 193, by 
the Alexander von Humboldt Stiftung, 
and by the Croatian Ministry of Science - project 0119250.

\newpage

\begin {table}[]
\begin {center}
\caption {The parameter set PC-F1 from Ref.\protect\cite{BMM.02}.}
\begin {tabular}{ccc}
\\ \hline 
Coupling constant  & Dimension  &  Value   \\ \hline
$\alpha_S$        &  MeV$^{-2}$  &  $-3.83577\times 10^{-4}$\\
$\beta_S$         &  MeV$^{-5}$  &  $7.68567\times 10^{-11}$ \\
$\gamma_S$        &  MeV$^{-8}$  &  $-2.90443\times 10^{-17}$ \\
$\delta_S$        &  MeV$^{-4}$  &  $-4.1853\times 10^{-10}$ \\
$\alpha_V$        &  MeV$^{-2}$  &  $2.59333\times 10^{-4}$  \\
$\gamma_V$        &  MeV$^{-8}$  &  $-3.879\times 10^{-18}$  \\
$\delta_V$        &  MeV$^{-4}$  &  $-1.1921\times 10^{-10}$ \\
$\alpha_{TV}$     &  MeV$^{-2}$  &  $3.4677\times 10^{-5}$ \\
$\delta_{TV}$     &  MeV$^{-4}$  &  $-4.2\times 10^{-11}$ \\

\end {tabular}
\label{tab0}
\end{center}
\end{table}

\begin {table}[]
\begin {center}
\caption {The R(Q)RPA excitation energies of the ISGMR, calculated with
   the PC-F1 effective interaction.
   The theoretical $m_1 /m_0$ centroids are compared with the experimental 
   excitation
   energies of the monopole resonances from Refs.\protect\cite{Young1}
   ($^{90}$Zr), \protect\cite{Young2} ($^{116}$Sn, $^{144}$Sm, $^{208}$Pb), and
   \protect\cite{Young3} ($^{112}$Sn, $^{124}$Sn).}
\begin {tabular}{ccc}
\\ \hline 
            & PC-F1  (MeV)      &  EXP   (MeV) \\ \hline
$^{90}$Zr   &  $18.6$   &  $17.81 + 0.12 - 0.12$   \\
$^{112}$Sn  &  $17.0$   &  $15.43 + 0.11 - 0.10 $ \\
$^{116}$Sn  &  $17.0$   &  $15.82 + 0.20 - 0.20$ \\
$^{124}$Sn  &  $16.5$   &  $14.50 + 0.14 - 0.14$ \\
$^{144}$Sm  &  $16.0$   &  $15.40 + 0.30 - 0.30$  \\
$^{208}$Pb  &  $14.2$   &  $13.96 + 0.20 - 0.20$  \\
\end {tabular}
\label{tab1}
\end{center}
\end{table}

\begin {table}[]
\begin {center}
\caption {The R(Q)RPA excitation energies of the IVGDR in 
$^{90}$Zr, $^{116}$Sn,
 $^{118}$Sn, $^{120}$Sn, and $^{124}$Sn, calculated with the PC-F1 effective
 interaction. The theoretical centroids are compared with the experimental data
 from Ref.\protect\cite{BF.75}. The centroid energy $E_{GDR}=m_1/m_0$ is
 calculated in the same energy window as the one used in the experimental
 analysis ($14-19$ MeV for the $^{90}$Zr, and $13-18$ MeV for the tin isotopes).}
\begin {tabular}{ccc}
\\ \hline
            & PC-F1 (MeV)   &  EXP (MeV)   \\ \hline
$^{90}$Zr   &  $16.22$          &  $16.85$ \\
$^{116}$Sn  &  $15.34$          &  $15.68$   \\
$^{118}$Sn  &  $15.25$          &  $15.59$  \\
$^{120}$Sn  &  $15.09$          &  $15.40$   \\
$^{124}$Sn  &  $15.03$          &  $15.29$  \\
\end {tabular}
\label{tab2}
\end{center}
\end{table}

\begin {table}[]
\begin {center}
\caption {The R(Q)RPA excitation energies of the ISGQR, calculated with 
   the PC-F1 effective interaction.
   The theoretical $m_1 /m_0$ centroids are compared with the experimental 
   excitation
   energies of the quadrupole resonances from Refs.\protect\cite{Young1}
   ($^{90}$Zr), \protect\cite{Young2} ($^{116}$Sn, $^{144}$Sm, $^{208}$Pb), and
   \protect\cite{Young3} ($^{112}$Sn, $^{124}$Sn).}
\begin {tabular}{ccc}
\\ \hline
            & PC-F1 (MeV)      &  EXP (MeV)   \\ \hline
$^{90}$Zr   &  $15.9$          &  $14.30 + 0.4 - 0.12$ \\
$^{112}$Sn  &  $15.3$          &  $13.23 +0.18 - 0.14$   \\
$^{116}$Sn  &  $14.8$          &  $13.30 + 0.35 - 0.35$   \\
$^{124}$Sn  &  $14.6$          &  $12.81 + 0.14 - 0.10$   \\
$^{144}$Sm  &  $14.2$          &  $12.78 + 0.30 - 0.30$  \\
$^{208}$Pb  &  $12.1$          &  $10.89 + 0.30 - 0.30$     \\
\end {tabular}
\label{tab3}
\end{center}
\end{table}


\bigskip

\begin{thebibliography}{999}
\bibitem{BHR.03} M. Bender, P.-H. Heenen, and P.-G. Reinhard,
	Rev. Mod. Phys. 75, 121 (2003).
	
\bibitem{VALR.05} D. Vretenar, A.V. Afanasjev, G.A. Lalazissis, and 
	P. Ring, Phys. Rep. 409, 101 (2005).
	
\bibitem{BB.77}
J. Boguta and A. R. Bodmer, Nucl. Phys. A 292, 413 (1977).

\bibitem{LKR.97}
G. A. Lalazissis, J. K{\"o}nig, and P. Ring, Phys. Rev. C 55, 540 (1997).

\bibitem{FLW.95} C. Fuchs, H. Lenske, and H.H. Wolter,
        Phys. Rev. C 52, 3043 (1995).
	
\bibitem{JL.98} F. de Jong and H. Lenske,
     Phys. Rev. C 57, 3099 (1998).

\bibitem{TW.99} S. Typel and H.H. Wolter,
     Nucl. Phys. A 656, 331 (1999).

\bibitem{NVF.02}
	T. Nik{\v{s}}i{\'{c}}, D. Vretenar, P. Finelli, and P. Ring, 
	Phys. Rev. C 66,  024306  (2002).

\bibitem{MM.89}
	P. Manakos and T. Mannel, Z. Phys. A 334,  481  (1989).

\bibitem{MNH.92}
	D. G. Madland, B. A. Nikolaus, and T. Hoch, 
	Phys. Rev. C 46, 1757 (1992).
	
\bibitem{Hoch.94} T. Hoch, D. Madland, P. Manakos, T. Mannel, B.A. Nikolaus,
      and D. Strottman, Phys. Rep. 242, 253 (1994).

\bibitem{FML.96}
	J. L. Friar, D. G. Madland, and B. W. Lynn, 
	Phys. Rev. C 53, 3085 (1996).

\bibitem{RF.97}
	J. J. Rusnak and R. J. Furnstahl, 
	Nucl. Phys. A 627,  495  (1997).

\bibitem{BMM.02}
	T. B{\"u}rvenich, D. G. Madland, J. A. Maruhn, and P.-G. Reinhard, 
	Phys. Rev. C 65,  044308  (2002).
	
\bibitem{FKV.03}
	P. Finelli, N. Kaiser, D. Vretenar, and W. Weise, 
	Eur. Phys. J. A 17,  573 (2003).
	
\bibitem{FKV.04}
	P. Finelli, N. Kaiser, D. Vretenar, and W. Weise, 
	Nucl. Phys. A 735, 449 (2004).
	
\bibitem{VW.04}
D. Vretenar and W. Weise,  in Lecture Notes in Physics, Vol.~641, edited by G.
  Lalazissis, P. Ring, and D. Vretenar (Springer-Verlag, Heidelberg, 2004).
  
\bibitem{FS.00b} R.J. Furnstahl and B.D. Serot,
        Nucl. Phys. A{\bf 671}, 447 (2000).
	 
\bibitem{BMR.04} T. B{\"u}rvenich, D. G. Madland, and P.-G. Reinhard,
	Nucl. Phys. A 744, 92 (2004).
	
\bibitem{NVR.02} T. Nik{\v{s}}i{\'{c}}, D. Vretenar, and P. Ring, 
	Phys. Rev. C 66, 064302 (2002).

\bibitem{LNV.05} G.A. Lalazissis, T. Nik\v si\' c, D. Vretenar, and P. Ring,
	Phys. Rev. C 71, 024312 (2005).
	
\bibitem{Ma.97} Z.Y. Ma, N. Van Giai, H. Toki, and M. L'Huillier, 
    Phys. Rev. C 55, 2385 (1997).
    
\bibitem{VWR.00} D. Vretenar, A. Wandelt, and P. Ring, 
    Phys. Lett. B 487, 334 (2000).
    
\bibitem{DF.90} J.F. Dawson and R.J. Furnstahl, 
    Phys. Rev. C 42, 2009 (1990).
    
\bibitem{Rin.01} P. Ring, Zhong-yu Ma, Nguyen Van Giai, D. Vretenar, 
	A. Wandelt, and Li-gang Cao, Nucl. Phys. A 694, 249 (2001).
	
\bibitem{Paar.03} N. Paar, P. Ring, T. Nik\v{s}i\'{c}, and D. Vretenar,
    Phys. Rev. C 67, 034312 (2003).

\bibitem{Ma.01} Z.Y. Ma, N. Van Giai, A. Wandelt, D. Vretenar, and P. Ring,
    Nucl. Phys. A 686, 173 (2001).
    
\bibitem{Pie.01} J. Piekarewicz, Phys. Rev. C 64, 024307 (2001).

\bibitem{Ma.02} Z.Y. Ma, A. Wandelt, N. Van Giai, D. Vretenar, P. Ring, and
    L.G. Cao, Nucl. Phys. A 703, 222 (2002).
    
\bibitem{VPRLa.01} D. Vretenar, N. Paar, P. Ring, and G.A. Lalazissis,
    Phys. Rev. C 63, 047301 (2001).
    
\bibitem{VPRLb.01} D. Vretenar, N. Paar, P. Ring, and G.A. Lalazissis, 
    Nucl. Phys. A 692, 496 (2001). 
    
\bibitem{VPNR.02} D. Vretenar, N. Paar, T. Nik\v{s}i\'{c}, and P. Ring,
    Phys. Rev. C 65, 021301(R) (2002).

\bibitem{Paar.04} N. Paar, T. Nik\v{s}i\'{c}, D. Vretenar, and P. Ring, 
    Phys. Rev. C 69, 054303 (2004).	
    
\bibitem{VBR.95}  D. Vretenar, H. Berghammer, and P. Ring, 
	Nucl. Phys. A 581, 679 (1995).
	
\bibitem{BGG.84} J.F. Berger, M. Girod, and D. Gogny, 
	Nucl. Phys. A 428,  23c (1984).
	
\bibitem{BGG.91} J. F. Berger, M. Girod, and D. Gogny, 
	Commput. Phys. Commun. 63, 365 (1991).
	
\bibitem{Bla.80} J.P. Blaizot, Phys. Rep. 64, 171 (1980).

\bibitem{BBDG.95} J.P. Blaizot, J.F. Berger, J. Decharg\' e, and M. Girod,
    Nucl. Phys. A 591, 435 (1995).
    
\bibitem{VNR.03} D. Vretenar, T. Nik\v{s}i\'{c}, and P. Ring, 
    Phys. Rev. C 68, 024310 (2003).
    
\bibitem{Young2} D.H. Youngblood, Y.-W. Lui, H.L. Clark, B. John, Y. Tokimoto,
               and X. Chen, Phys. Rev. C 69, 034315 (2004).
	       
\bibitem{Young1} D.H. Youngblood, Y.-W. Lui, B. John, Y. Tokimoto, H.L. Clark,
               and X. Chen, Phys. Rev. C 69, 054312 (2004).	       
       
\bibitem{Young3} Y.-W. Lui, D.H. Youngblood, Y. Tokimoto, H.L. Clark, 
               and B. John, Phys. Rev. C 70, 014307 (2004).
	       	       	     
\bibitem{Lee.98} C.-H. Lee, T.T.S. Kuo, G.Q. Li, and G.E. Brown, 
    Phys. Rev. C 57, 3488 (1998).
    
\bibitem{Rit.93} J. Ritman et al., Phys. Rev. Lett 70, 533 (1993).

\bibitem{BF.75} B.L. Berman and S.C. Fultz, 
         Rev. Mod. Phys. 47, 713(1975). 

\bibitem{Rei.99} P.-G. Reinhard, Nucl. Phys. A 649, 305c (1999).

\bibitem{JM.89} M. Jaminon and C. Mahaux, Phys. Rev. C 40, 354 (1989).

\bibitem{JM.90} M. Jaminon and C. Mahaux, Phys. Rev. C 41, 697 (1990).

\bibitem{VNR.02} D. Vretenar, T. Nik\v si\' c, and P. Ring,
	Phys. Rev. C 65, 024321 (2002).
		        	  
         	   
 
\end{thebibliography}
\end{document}